\documentclass{aa}
\begin{document}
\newcommand\bsec{\hbox{$.\!\!{\arcsec}$}}
\newcommand\rsec{\hbox{$.\!\!{^s}$}}
\newcommand\RA[4]{#1$^{\rm h}$#2$^{\rm m}$#3\rsec#4}
\newcommand\DEC[3]{#1$^{\circ}$#2\arcmin#3\arcsec}
\newcommand\teff{$T_{\rm eff}$} 
\newcommand\logt{$ \log\,T_{\rm eff}$}
\newcommand\logg{$ \log\,g$}
\newcommand\loghe{$ \log{\frac{n_{\rm He}}{n_{\rm H}}}$}
\newcommand\Msolar{$M_\odot$}
\thesaurus{01 (08.05.1; 08.06.3; 08.08.2; 10.07.3 NGC~6752)}
\subtitle{{\sl Letter to the Editor}\\}
\title{Physical Parameters of Hot Horizontal-Branch Stars in NGC 6752: Deep 
Mixing and Radiative Levitation}
\author{S.~Moehler\inst{1}\fnmsep\thanks{Based on observations collected at the 
 European Southern Observatory (ESO~N$^{\b{o}}$~60.E-145, 61.E-0145, 
61.E-0361)} 
 \and A.~V.~Sweigart\inst{2} 
 \and W.B.~Landsman\inst{3} \and 
 U.~Heber\inst{1}
 \and M.~Catelan\inst{2}\fnmsep\thanks{Hubble Fellow.}\fnmsep\thanks{Visiting
 Scientist, Universities Space Research Association.}}
\offprints{S.~Moehler}
\institute {Dr. Remeis-Sternwarte, Astronomisches Institut der Universit\"at
 Erlangen-N\"urnberg, Sternwartstr. 7, 96049 Bamberg, Germany\\
 e-mail: moehler@sternwarte.uni-erlangen.de, heber@sternwarte.uni-erlangen.de
 \and NASA\,Goddard Space Flight Center, Code 681, Greenbelt, 
 MD 20771, USA \\
 e-mail: sweigart@bach.gsfc.nasa.gov, catelan@stars.gsfc.nasa.gov
 \and Raytheon ITSS, NASA\,Goddard Space Flight Center, Code 681, Greenbelt,
 MD 20771, USA \\
 e-mail: landsman@mpb.gsfc.nasa.gov
}
\date{}
\titlerunning{Hot HB stars in NGC~6752 - Deep mixing and radiative 
levitation}
\maketitle

\begin{abstract}

Atmospheric parameters (\teff, \logg\ and \loghe ) are derived for 42 hot 
horizontal branch (HB) stars in the globular cluster NGC\,6752. For 19 stars 
\ion{Mg}{II} and \ion{Fe}{II} lines are detected indicating an iron enrichment
by a factor 50 on average with respect to the cluster abundance whereas the
magnesium abundances are consistent with the cluster metallicity. This finding
adds to the growing evidence that radiative levitation plays a significant
role in determining the physical parameters of blue HB stars. Indeed, we find
that iron enrichment can explain part, but not all, of the problem of
anomalously low gravities along the blue HB. Thus the physical parameters of
horizontal branch stars hotter than about $11,\!500$~K in NGC~6752, as derived
in this paper, are best explained by a combination of helium mixing and
radiative levitation effects. 

\end{abstract}


\section{Introduction} 

The discovery of ``gaps'' along the blue horizontal branch (HB) in globular 
clusters as well as of long extensions towards higher temperatures has
triggered several spectroscopic investigations (Moehler \cite{moeh99} and
references therein) yielding the following results:

\begin{enumerate}
\item

Most of the stars analysed above and below any gaps along the blue horizontal
branch are ``bona fide'' blue HB stars ($T_{\rm eff} < 20,000$~K), which show
significantly lower gravities than expected from canonical stellar evolution
theory.

\item

Only in NGC~6752 and M~15 have spectroscopic analyses verified the presence of
stars that could be identified with the subdwarf B stars known in the field of
the Milky Way ($T_{\rm eff} > 20,000$~K, \logg\ $>$ 5). In contrast to the
cooler blue HB (BHB) stars the gravities of these ``extended HB'' (EHB) stars 
agree well with the expectations of canonical stellar evolution.

\end{enumerate}

Two scenarios have been suggested to account for the low gravities of BHB
stars:

\begin{description}
\item [{\bf Helium mixing scenario:}]

Abundance anomalies observed in red giant branch (RGB) stars in globular 
clusters (e.g., Kraft \cite{kraf94}, Kraft et al. \cite{krsn97}) may be
explained by the dredge-up of nuclearly processed material to the stellar
surface. If the mixing currents extend into the hydrogen-burning shell -- as
suggested by current RGB nucleosynthesis models and observed Al overabundances
-- helium can be mixed into the stellar envelope. This in turn would increase
the luminosity (and mass loss) along the RGB (Sweigart \cite{swei99}) and
thereby create less massive (i.e. bluer) HB stars with helium-enriched hydrogen
envelopes. The helium enrichment increases the hydrogen burning rate, leading
to higher luminosities (compared to canonical HB stars of the same temperature)
and lower gravities. The gravities of stars hotter than about 20,000~K are not
affected by this mixing process because these stars have only inert hydrogen
shells. 

\item [{\bf Radiative levitation scenario:}]

Grundahl et al. (\cite{grca99}) found a ``jump'' in the $u$, $u-y$ 
colour-magnitude diagrams of 15 globular clusters, which can be explained if
radiative levitation of iron and other heavy elements takes place over the
temperature range defined by the ``low-gravity'' BHB stars. This assumption has
been confirmed in the case of M~13 by the recent high resolution spectroscopy
of Behr et al. (\cite{beco99}). Grundahl et al. argue that super-solar
abundances of heavy elements such as iron should lead to changes in model
atmospheres which may be capable of explaining the disagreement between models
and observations over the ``critical'' temperature range $11,\!500~{\rm K} <
T_{\rm eff} < 20,\!000$~K.

\end{description}

NGC~6752 is an ideal test case for these scenarios, since it is a very close
globular cluster with a long blue HB extending to rather hot EHB stars. While
previous data already cover the faint end of the EHB, we now obtained new
spectra for 32 stars in and above the sparsely populated region between the 
BHB and the EHB stars. In this {\em Letter}, we present atmospheric parameters 
derived for a total of 42 BHB and EHB stars and discuss the constraints they 
may pose on the scenarios described above.

\section{Observational Data}

\begin{figure}
\vspace{7.cm}
\includegraphics{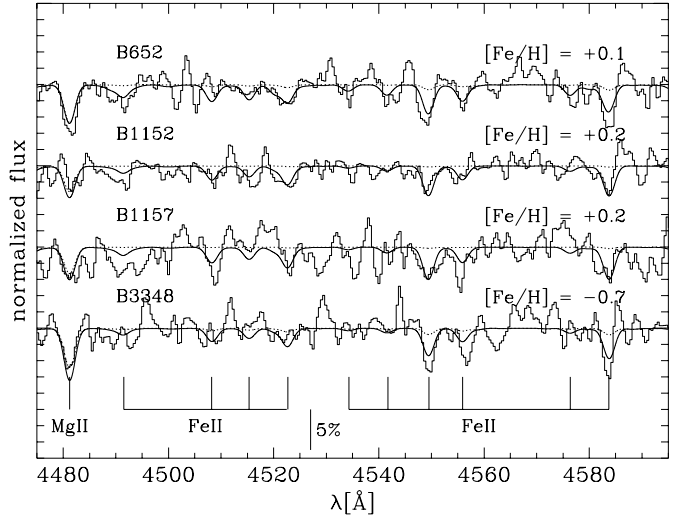}
\caption{The iron and magnesium lines as seen in the spectra of some of the 
stars in NGC~6752. The solid line marks a model spectrum for which we used a 
solar metallicity model stratification but adjusted all metals to [M/H] = 
$-$1.5 for the spectrum synthesis except iron. The iron abundance was then
adjusted to the noted values to reproduce the marked \ion{Fe}{II} lines. The
dashed line marks a model spectrum for which we used the cluster metal
abundance for all metals. Effective temperature and surface gravity for both
models are those plotted in Fig.~2, upper and central panel, respectively. The
temperatures of the stars range from 12,000~K (B3348) to 16,000~K (B1152).
Obviously iron is strongly enriched whereas magnesium is consistent with the
mean cluster abundance.\label{n6752bhb_speciron}} 

\end{figure}

We selected our targets from the photographic photometry of Buonanno et al.
(\cite{buca86}) to cover the range 14.5$\le V \le$15.5. 19 stars were observed
with the ESO 1.52m telescope (61.E-0145, July 22-25, 1998) and the Boller \&
Chivens spectrograph using CCD \# 39 and grating \# 33 (65~\AA/mm). This
combination covered the 3300~\AA\ -- 5300~\AA\ region at a spectral resolution 
of 2.6~\AA . The data reduction will be described in Moehler et al.
(\cite{mosw99}). Prompted by the suggestion of Grundahl et al. (\cite{grca99})
that radiative levitation of heavy metals may enrich the atmospheres of BHB
stars, we searched for metal absorption lines in these spectra. Indeed we found
\ion{Fe}{II} absorption lines in almost all spectra (for examples see
Fig.~\ref{n6752bhb_speciron}).

13 stars were observed as backup targets at the NTT during observing runs
dedicated to other programs (60.E-0145, 61.E-0361). The observations and their
reduction are described in Moehler et al. (\cite{mola99}). Those spectra have a
spectral resolution of 5~\AA\ covering 3350 to 5250~\AA. No metal lines could
be detected due to this rather low spectral resolution.

\section{Atmospheric Parameters}

\subsection{Fit procedure and model atmospheres}

To derive effective temperatures, surface gravities and helium abundances we
fitted the observed Balmer lines H$_\beta$ to H$_{10}$ (excluding H$_\epsilon$ 
because of possible blending problems with the \ion{Ca}{II} H line) and the 
helium lines (\ion{He}{I} 4026, 4388, 4471, 4922\AA) with stellar model
atmospheres. We corrected the spectra for radial velocity shifts, derived from
the positions of the Balmer and helium lines and normalized the spectra by eye.

We computed model atmospheres using ATLAS9 (Kurucz \cite{kuru91}) and used 
Lemke's version of the LINFOR\footnote{For a description see 
http://a400.sternwarte.uni-erlangen.de/$\sim$ai26/linfit/linfor.html} program
(developed originally by Holweger, Steffen, and Steenbock at Kiel University)
to compute a grid of theoretical spectra which include the Balmer lines
H$_\alpha$ to H$_{22}$ and \ion{He}{I} lines. The grid covered the range
7,000~K~$\leq$~\teff~$\leq$~35,000~K, 2.5~$\leq$~\logg~$\leq$~5.0, 
$-3.0$~$\leq$~\loghe~$\leq$~$-1.0$, at a metallicity of [M/H]~=~$-1.5$. 

\begin{figure}
\vspace{11.5cm}
\includegraphics{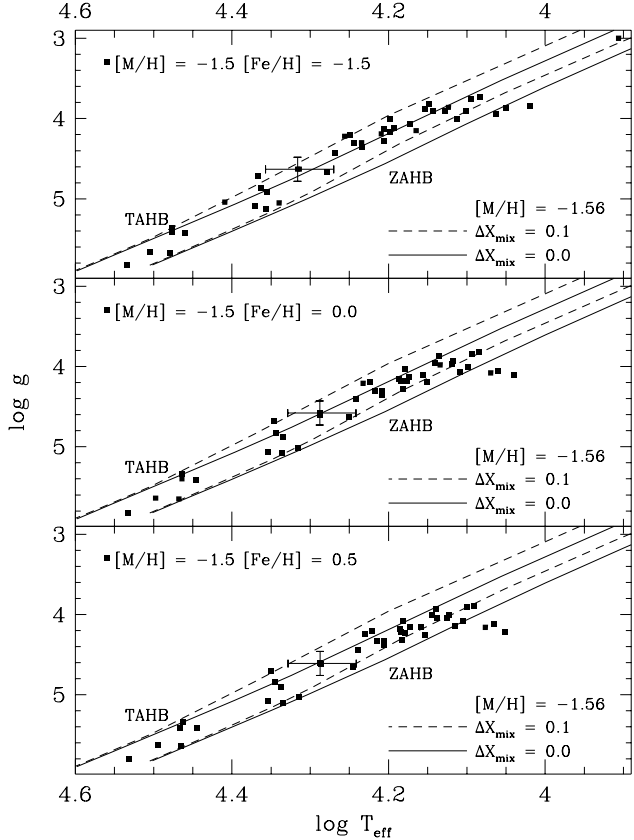}
\caption{Temperatures and gravities of the programme stars in 
NGC~6752. 
{\bf upper panel}: determined from models with cluster metallicity 
([M/H] = $-$1.5), 
{\bf central panel}: adopting a solar metallicity model stratification 
([M/H] = 0) and spectrum synthesis with solar iron abundance but cluster 
abundances for all other metals M ([M/H]=$-$1.5)
{\bf lower panel}: adopting a super--solar metallicity model stratification 
([M/H] = $+$0.5) and iron abundance ([Fe/H] = $+$0.5) but cluster abundances 
([M/H] = $-$1.5) for all other metals in the spectrum synthesis.
For more details see text.
Also plotted are the zero-age HB (ZAHB) and terminal-age HB (TAHB, i.e., 
central helium exhaustion) from the Sweigart (1999) tracks for metallicity 
[M/H] = -1.56. The dashed and solid lines correspond to tracks with and without
 mixing, respectively. $\Delta X_{\rm mix}$ measures the difference in 
hydrogen abundance $X$ between the envelope ($X = X_{\rm env}$) and the 
innermost point reached by the mixing currents 
($X = X_{\rm env}-\Delta X_{\rm mix}$) in the red giant precursors 
and is thus an indicator for the amount of helium mixed into the envelope of 
the red giant. Representative error bars are plotted\label{n6752bhb_irontg}} 

\end{figure}

To establish the best fit we used the routines developed by Bergeron et al.
(\cite{besa92}) and Saffer et al. (\cite{saff94}), which employ a $\chi^2$ 
test. The fit program normalizes model spectra {\em and} observed spectra using
the same points for the continuum definition.
The results are plotted in Fig.~\ref{n6752bhb_irontg} (upper panel). The errors
are estimated to be about 10\% in \teff\ and 0.15~dex in \logg\ (cf. Moehler et
al. \cite{mohe97}). Representative error bars are shown in
Fig.~\ref{n6752bhb_irontg} . To increase our data sample we reanalysed the NTT
spectra described and analysed by Moehler et al. (\cite{mohe97}). For a
detailed comparison see Moehler et al. (\cite{mosw99}). 

\subsection{Iron abundances}

Due to the spectral resolution and the weakness of the few observed lines a
detailed abundance analysis (such as that of Behr et al., 1999) is beyond the
scope of this paper. Nevertheless we can estimate the iron abundance in the
stars by fitting the \ion{Fe}{II} lines marked in Fig.~\ref{n6752bhb_speciron}.
A first check indicated that the iron abundance was about solar whereas the
magnesium abundance was close to the mean cluster abundance.

As iron is very important for the temperature stratification of stellar
atmospheres we tried to take the increased iron abundance into account: We used
ATLAS9 to calculate a solar metallicity atmosphere. The emergent spectrum was
then computed from the solar metallicity model stratification by reducing the
abundances of all metals M (except iron) to the cluster abundances ([M/H] =
$-$1.5). It was not possible to compute an emergent spectrum that was fully
consistent with this iron-enriched composition, since the ATLAS9 code requires
a scaled solar composition. We next repeated the fit to derive \teff , \logg,
and \loghe\ with these enriched model atmospheres. The results are plotted in 
Fig.~\ref{n6752bhb_irontg} (central panel).

\begin{figure}
\vspace{12cm}
\includegraphics{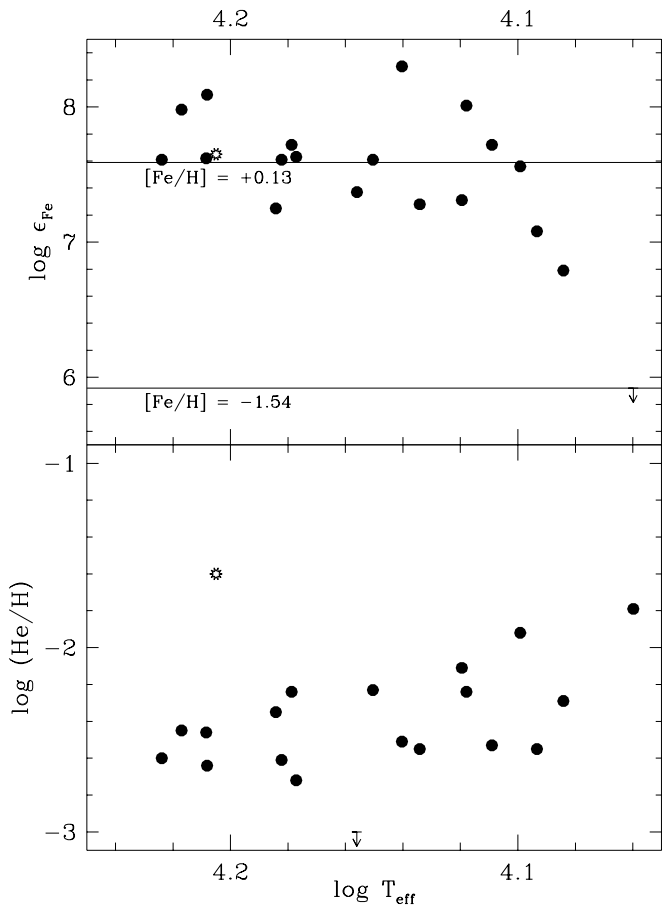}
\caption{The iron and helium abundances for the stars observed with ESO 1.52m
telescope. Iron was not detected in the coolest star and is plotted as an 
upper limit. The trend to lower helium abundances for higher temperatures
agrees with the findings of Behr et al. (1999). Iron is obviously enhanced to
roughly solar abundances. The mean iron abundance as derived from our spectra
([Fe/H] = $+$0.13) and the cluster abundance ([Fe/H] = $-$1.54) are marked. The
asterisk marks the results of Glaspey et al. (1989) for the hotter of their two
BHB stars in NGC~6752\label{n6752bhb_iron}} 
\end{figure}

For each star observed at the ESO 1.52m telescope we then computed an
``iron-enriched'' model spectrum with \teff , \logg\ as derived from the fits
of the Balmer and helium lines with the ``enriched'' model atmospheres (cf.
Fig.~\ref{n6752bhb_irontg}, central panel) and \loghe\ = $-$2. The fit of the
iron lines was started with a solar iron abundance and the iron abundance was
varied until $\chi^2$ achieved a minimum. As the radiative levitation in BHB
stars is due to diffusion processes (which is also indicated by the helium
deficiency found in these stars) the atmospheres have to be very stable. We
therefore kept the microturbulent velocity $\xi$ at 0\,km/s -- the iron
abundances plotted in Fig.~\ref{n6752bhb_iron} are thus upper limits. The mean
iron abundance turns out to be [Fe/H] $\approx +0.1$~dex (for 18 stars hotter
than about 11,500~K) and $\le-$1.6 for the one star cooler than 11,500~K.
Although the iron abundance for the hotter BHB stars is about a factor of 50
larger than the cluster abundance, it is smaller by a factor of 3 than the
value of [Fe/H] = $+$0.5 estimated by Grundahl et al. (\cite{grca99}) as being
necessary to explain the Str\"omgren $u$-jump observed in $u$, $u-y$
colour-magnitude diagrams.

Our results are in good agreement with the findings of Behr et al.
(\cite{beco99}) for BHB stars in M~13 and Glaspey et al. (\cite{glmi89}) for
two BHB stars in NGC~6752. Again in agreement with Behr et al. (\cite{beco99})
we see a decrease in helium abundance with increasing temperature, whereas the
iron abundance stays roughly constant over the observed temperature range.

\subsection{Influence of iron enrichment}

From Fig.~\ref{n6752bhb_irontg} it is clear that the use of enriched model
atmospheres moves most stars closer to the zero-age horizontal branch (ZAHB).
The three stars between 10,000~K and 12,000~K, however, fall {\em below} the
canonical ZAHB when fitted with enriched model atmospheres. This is plausible
as the radiative levitation is supposed to start around 11,500~K (Grundahl et
al. \cite{grca99}) and the cooler stars therefore should have metal-poor
atmospheres (see also Fig.~\ref{n6752bhb_iron} where the coolest analysed star
shows no evidence of iron enrichment). We repeated the experiment by increasing
the iron abundance to [Fe/H]=$+$0.5 (see Fig.~\ref{n6752bhb_irontg} lower
panel), which did not change the resulting values for \teff\ and \logg\
significantly. 

Since HB stars at these temperatures spend most of their lifetime close to the
ZAHB, one would expect the majority of the stars to scatter (within the
observational error limits) around the ZAHB line in the \logt , \logg--diagram. 
However, this is not the case for the canonical ZAHB (solid lines in 
Fig.~\ref{n6752bhb_irontg}) even with the use of iron-enriched model
atmospheres (central and lower panels in Fig.~\ref{n6752bhb_irontg}). The
scatter instead seems more consistent with the ZAHB for moderate helium mixing
(dashed lines in Fig.~\ref{n6752bhb_irontg}). Thus the physical parameters of
HB stars hotter than $\approx 11,\!500$~K in NGC~6752, as derived in this
paper, are best explained by a combination of helium mixing and radiative
levitation effects.

\section{Conclusions}

Our conclusions can be summarized as follows:

\begin{enumerate}

\item 

We have obtained new optical spectra of 32 hot HB stars in NGC 6752 with 11,000
K $<$ \teff $<$ 25,000. When these spectra (together with older spectra of
hotter stars) are analysed using model atmospheres with the cluster metallicity
([Fe/H] = $-$1.5), they show the same ``low-gravity'' anomaly with respect to
canonical HB models, that has been observed in several other clusters (Moehler
1999).

\item 

For 18 stars with \teff\ $>$ 11,500 K, we estimate a mean iron abundance of
[Fe/H] $\approx +$0.1, whereas magnesium is consistent with the cluster 
metallicity. The hot HB stars in NGC 6752 thus show an abundance pattern 
similar to that observed in M~13 (Behr et al. \cite{beco99}), which presumably 
arises from radiative levitation of iron (Grundahl et al. 1999).

\item 

When the hot HB stars are analysed using model atmospheres with an
appropriately high iron abundance, the size of the gravity anomaly with 
respect to canonical HB models is significantly reduced. Whether the remaining
differences between observations and canonical theory can be attributed to
levitation effects on elements other than iron remains to be investigated by
detailed modeling of the diffusion processes in the stellar atmospheres. With
presently available models, the derived gravities for HB stars hotter than
$\approx 11,\!500$~K are best fit by non-canonical HB models which include deep
mixing of helium on the RGB (Sweigart \cite{swei99}). 

\end{enumerate}
\acknowledgements

We thank the staff of the ESO La Silla observatory for their support 
during our observations. S.M. acknowledges financial support from the DARA
under grant 50~OR~96029-ZA. M.C. was supported by NASA through
Hubble Fellowship grant HF--01105.01--98A awarded by the Space Telescope 
Science Institute, which is 
operated by the Association of Universities for Research in Astronomy, 
Inc., for NASA under contract NAS~5-26555. We are grateful to the referee, 
Dr. R. Kraft, for his speedy report and valuable remarks.

{}
\end{document}